# Revisiting UTAUT for the Age of AI: Understanding Employees' AI Adoption and Usage Patterns Through an Extended UTAUT Framework

Diana Wolfe[1], Matt Price[2], Alice Choe[3], Fergus Kidd[4], Hannah Wagner[5]

[1] Doctoral Candidate, Industrial Organizational Psychology, Seattle Pacific University, 3307 3rdAve. W, Seattle, WA

[2] PhD in Media Psychology & Behavioral Neuropsychology; AI Governance, Responsible AI and Emerging Technology Platform Lead, Avanade

[3] Doctoral Candidate, Organizational Behavior and Human Resource Management, Rotman School of Management, University of Toronto, 105 St. George Street, Toronto, Ontario, Canada

[4] Head of Engineering, Avanade, 30 Cannon Street, London, UK

[5] Doctoral Candidate, Industrial Organizational Psychology, Seattle Pacific University, 3307 3rdAve. W, Seattle, WA

Corresponding concerning this article should be addressed to: diana.wolfe@avanade.com

**Abstract**

This study investigates whether demographic factors shape employees' adoption and attitudes toward artificial intelligence (AI) technologies at work. Building on an extended Unified Theory of Acceptance and Use of Technology (UTAUT), which reintroduces affective dimensions such as attitude, self-efficacy, and anxiety, we surveyed 2,257 professionals across global regions and organizational levels within a multinational consulting firm. Non-parametric tests examined whether three demographic factors—years of experience, hierarchical level in the organization, and geographic region—were associated with AI adoption, usage intensity, and eight UTAUT constructs. Organizational level significantly predicted AI adoption, with senior employees showing higher usage rates, while experience and region were unrelated to adoption. Among AI users ($n$ = 1,256), frequency and duration of use showed minimal demographic variation. However, omnibus tests revealed small but consistent group differences across several UTAUT constructs, particularly anxiety, performance expectancy, and behavioral intention, suggesting that emotional and cognitive responses to AI vary modestly across contexts. These findings highlight that demographic factors explain limited variance in AI acceptance but remain relevant for understanding contextual nuances in technology-related attitudes. The results underscore the need to integrate affective and organizational factors into models of technology acceptance to support equitable, confident, and sustainable engagement with AI in modern workplaces.

**Keywords:** artificial intelligence, technology acceptance, UTAUT, demographic differences, workplace anxiety, organizational psychology

**Employee AI Adoption and Usage**

The rapid integration of artificial intelligence (AI) technologies into workplace environments represents one of the most consequential technological shifts of the 21st century. Unlike prior waves of digital transformation, such as enterprise resource planning systems or robotic process automation, that primarily streamlined routine or manual tasks, AI technologies possess a unique capability to augment or potentially replace cognitive work traditionally performed by knowledge professionals (Daugherty & Wilson, 2024). This qualitative leap raises profound questions not only about technical implementation but also about the psychological and organizational dimensions of human work: Will employees embrace AI as a productivity-enhancing partner, or resist it as a threat to their professional identity and relevance? How do individual differences shape these responses? And what can organizations do to facilitate successful adoption while maintaining employee trust and well-being?

As organizations deploy AI at unprecedented speed and scale, a fundamental tension has emerged. Global AI spending is projected to reach nearly $1.5 trillion in 2025 (Gartner, 2025), with approximately 95% of firms now experimenting with or using generative AI technologies (Bain, 2023). However, this rapid organizational investment has outpaced employee readiness. Survey data reveal a disconnect between these ambitions and employee readiness. For example, Bain & Company's generative AI readiness survey finds that roughly 95% of firms report experimenting with or using generative AI, whereas only a fraction of employees feel their organizations are ready to implement them effectively (Qlik, 2024).

Employees moreover report mixed emotions regarding AI adoption, expressing excitement about efficiency gains coupled with apprehension about potential risks. For instance, a recent EY survey reported that 75% of U.S. workers worry AI could make jobs obsolete, and 65% feel anxious about AI replacing their own role (EY, 2023). The *2024 Stanford AI Index Report* documents a sharp uptick in apprehension, with a majority of respondents—roughly 52% worldwide—reporting discomfort with AI systems, up more than 13 points in just two years (Stanford, 2024). In the U.S., a 2025 *Pew Research Center* study found that more than half of workers (52%) are worried about the future impact of AI use in the workplace; 32% think it will lead to fewer job opportunities for them in the long run; 33% feel overwhelmed by AI; and 36% feel hopeful about AI's workplace use, demonstrating the mixed emotions. These divergent reactions underscore that AI adoption is not merely a technical rollout but a psychological and cultural transformation that depends fundamentally on whether individual employees accept and effectively utilize these technologies in their daily work.

The stakes of this adoption challenge are considerable. Failed AI implementations not only represent wasted financial investments but can also erode employee trust, increase turnover, and create lasting resistance to future technological changes (Brougham & Haar, 2018). Research demonstrates that poorly executed AI initiatives can significantly impact employee loyalty and morale, particularly when implementation occurs without adequate consultation or support (Bhargava et al., 2021).

Conversely, successful AI adoption can enhance job satisfaction, improve performance, and create competitive advantages (McElheran et al., 2021; Brynjolfsson et al., 2021; Kim, 2022; Coombs et al., 2010). Organizations that effectively implement AI have achieved 1.5 times higher revenue growth, 1.6 times greater shareholder returns, and notably higher employee satisfaction scores (BCG, 2024). While AI adoption outcomes clearly depend on multiple levels of analysis such as organizational strategy and market conditions, individual-level factors represent a critical and underexamined determinant of success. Ultimately, AI adoption is as much a socio-cultural transformation as a technical rollout, and organizations must therefore understand individual employees' willingness to accept and effectively utilize these technologies in their daily work.

**Purpose of the Present Study**

Despite the practical urgency of understanding AI acceptance, research on individual differences in AI technology adoption remains limited. Existing studies have documented widespread anxiety about AI (Kim, Soh, et al., 2024) and identified general barriers to adoption (Dwivedi et al., 2021), but few have systematically examined how demographic factors moderate acceptance attitudes within active AI-using populations. This gap is particularly problematic given that technology acceptance theories consistently demonstrate that demographic characteristics—such as experience, organizational position, and cultural context—can significantly moderate adoption behaviors (Venkatesh et al., 2003). Understanding these individual differences is essential for developing targeted implementation strategies that address the specific concerns and expectations of different employee groups.

individual differences is essential for developing targeted implementation strategies that address the specific concerns and expectations of different employee groups.

The present study addresses this gap by examining demographic differences in AI technology acceptance among 1,256 AI-using professionals within a multinational consulting organization. We employ the Unified Theory of Acceptance and Use of Technology (UTAUT) framework (Venkatesh et al., 2003), one of the most influential and widely validated models in technology acceptance research, to assess how years of professional experience, organizational level, and geographic region shape AI-related attitudes and behavioral intentions. Importantly, we extend the traditional UTAUT framework to incorporate anxiety as a central outcome variable, recognizing that emotional responses to AI represent a critical yet underexplored dimension of technology acceptance in contexts involving potentially threatening or opaque systems (Lee, Jones-Jang, Chung, Kim, & Choi, 2024).

The remainder of this paper proceeds as follows. Section 2 introduces the UTAUT theoretical framework and discusses its extension to AI adoption contexts. Section 3 develops specific hypotheses regarding how years of experience, organizational level, and geographic region influence AI acceptance. Section 4 describes our methodology, including sample characteristics, measurement instruments, and analytical approach. Section 5 presents results from our analyses of demographic differences in AI adoption and UTAUT dimensions. Section 6 discusses theoretical and practical implications, limitations, and directions for future research.

**Theoretical Framework**

*The Unified Theory of Acceptance and Use of Technology (UTAUT)*

The Unified Theory of Acceptance and Use of Technology (UTAUT) provides the primary theoretical foundation for this study. Developed by Venkatesh and colleagues (2003) to consolidate fragmented streams of technology acceptance research, the UTAUT synthesized key elements from eight prominent models—including the Theory of Reasoned Action (TRA), the Technology Acceptance Model (TAM), Innovation Diffusion Theory (IDT), Social Cognitive Theory (SCT), and others—into a single comprehensive theory. Over the past two decades, the UTAUT has become one of the most influential models in information systems, with hundreds of empirical applications across domains such as e-government services, mobile technologies, and enterprise systems.

At its core, UTAUT posits that four constructs shape individuals' behavioral intention to use technology and their subsequent usage behavior: performance expectancy, effort expectancy, social influence, and facilitating conditions. *Performance expectancy* refers to the perceived benefit or utility of the technology for improving one's job performance. *Effort expectancy* refers to perceived ease of use of the technology, capturing how simple or free of effort the technology is understood to be. *Social influence* refers to the extent to which an individual perceives that important others (e.g., colleagues, supervisors) believe they should use the new system. This construct is similar to subjective norm and tends to be especially salient in environments where use may be mandated or highly encouraged, though its effect weakens with experience (Venkatesh et al., 2003). Finally, *facilitating conditions* refer to the degree to which an individual believes that organizational and technical infrastructure exists to support use of the system. In other words, facilitating conditions reflect the availability of resources, training, and assistance necessary for effective use.

A distinguishing feature of UTAUT is its inclusion of key moderators that adjust the strength of these relationships. Venkatesh et al. (2003) identified four moderators: gender, age, experience, and voluntariness of use (see Figure 1). For example, they theorized that the influence of effort expectancy on intention is stronger for women and older individuals, but that it diminishes with increasing experience using the technology (Venkatesh et al., 2003). Similarly, social influence effects are more pronounced for women, older users, and in mandatory settings (low voluntariness), and they wane as users gain experience and confidence (Venkatesh et al., 2003). Empirically, voluntariness of use was specifically found to moderate the impact of social influence. When usage is mandated by the organization, pressure from others significantly drives intentions, whereas in purely voluntary contexts social influence carries less weight (Venkatesh et al., 2003). These nuanced moderating effects acknowledge that technology acceptance is not one-size-fits-all; individual differences and context can shape which factors matter most in a given situation.

**Figure 1**

*Unified Theory of Acceptance and Use of Technology (Source: Venkatesh et al., 2003)*

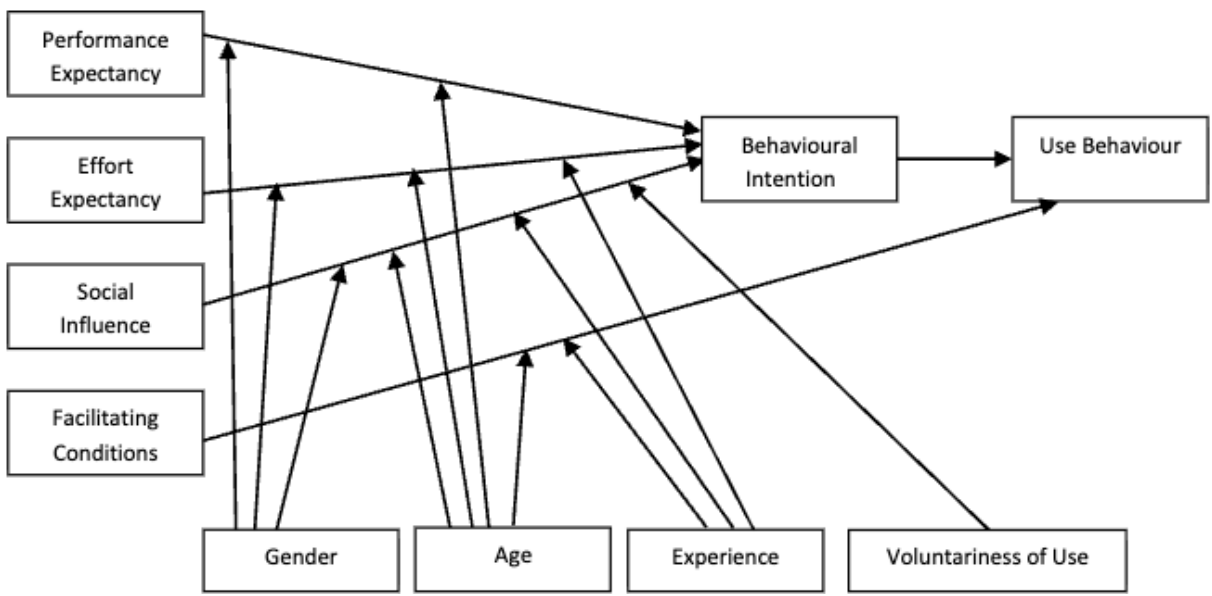

UTAUT theorized that performance expectancy, effort expectancy, and social influence all have direct positive effects on a user's behavioral intention to use a technology, while facilitating conditions influence actual usage behavior, especially when usage is not entirely voluntary (Venkatesh et al., 2003). Overall, the original validation study demonstrated that UTAUT explained up to 70% of the variance in behavioral intention (Venkatesh et al., 2003). Empirical tests of the model have borne out many of Venkatesh et al.'s (2003) theorized relationships, with performance expectancy emerging as a potent driver of intention fairly consistently (Marikyan & Papagiannidis, 2021; Venkatesh et al., 2003; Zhou et al., 2010; Abbad, 2021; Taiwo & Downe, 2013).

Effort expectancy also positively influences intention, though its effect tends to diminish with extended usage of technology (Gupta, Dasgupta, & Gupta, 2008; Ghauhan & Jaiswal, 2016); that is, once people become familiar with a system, ease of use becomes less of a differentiating factor. Social influence has been found to significantly affect intention primarily in the early stages of adoption or in mandatory-use contexts (Venkatesh et al., 2003). Its impact weakens, however, as users accumulate experience and form their own opinions about the technology (Venkatesh & Davis, 2000).

Empirical research largely supports Venkatesh et al.'s (2003) theorizing that facilitating conditions exert both indirect and direct effects on technology use. In early adoption stages, their influence on behavioral intention is largely absorbed by related beliefs about usefulness and ease of use, rendering their unique effect nonsignificant (Venkatesh et al., 2003; Williams et al., 2015). However, as users gain experience, facilitating conditions become a direct determinant of actual use, reflecting the enabling role of organizational and technical support in translating intention into behavior (Venkatesh et al., 2012; Dwivedi et al., 2019). Meta-analytic evidence confirms this pattern, showing robust effects of facilitating conditions on usage behavior and stronger associations among experienced users and in mandatory contexts (Blut et al., 2022).

Scholars have also noted important limitations of the original framework. UTAUT's emphasis is primarily cognitive, focusing on rational cost–benefit assessments while giving less weight to affective and contextual factors (Dwivedi et al., 2019). Subsequent research indicates that emotional and social–contextual variables, such as trust, anxiety, and perceived risk, can significantly shape technology adoption, particularly in settings where systems are opaque or usage is mandatory (Dwivedi et al., 2019; Williams et al., 2015). To address these gaps, Dwivedi and colleagues (2019) have advocated reintroducing previously excluded predictors, such as users' attitudes toward technology, to better capture the affective and experiential dimensions of technology acceptance (see Figure 2). In addition, empirical findings suggest that effort expectancy often becomes nonsignificant after extended use (Gupta et al., 2008; Chauhan & Jaiswal, 2016), implying a ceiling effect in experienced user samples.

**Figure 2**

*Proposed Adaptation of the UTAUT model (Source: Dwivedi et al., 2019)*

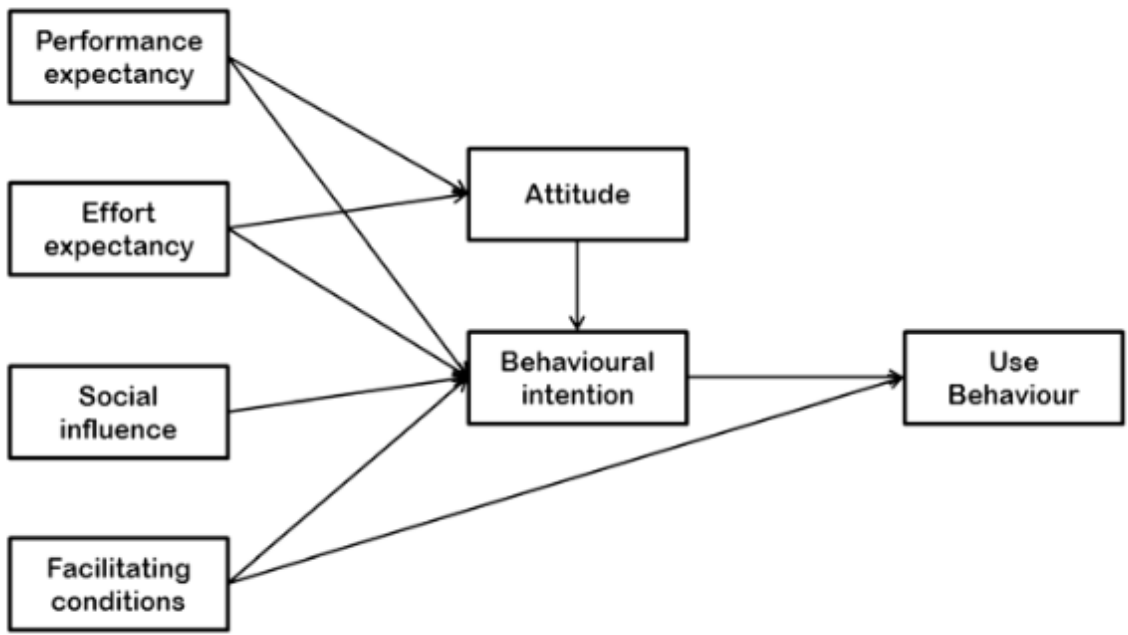

*Extension of the UTAUT to AI Adoption Context in Organizations*

Although the UTAUT framework has demonstrated strong predictive validity across traditional information system contexts, its focus remains primarily cognitive and utilitarian, emphasizing rational cost–benefit assessments of technology use. As Venkatesh and colleagues (2016) point out, one enduring weakness with the model and its application to different empirical contexts is its limited contextual theorizing. Most studies test UTAUT *in* context (e.g., education, e-government, mobile applications), but rarely theorize *about* context, such as how organizational culture and professionals norms shape technology acceptance dynamics. We argue that this narrow orientation is insufficient for understanding responses to artificial intelligence (AI) in professional workplaces—a class of technologies that evoke complex emotional, identity-based, and efficacy-related reactions that extend well beyond perceived usefulness or ease of use (Jones-Jang, Chung, et al., 2025; Sun, Xu, & Xu, 2024; McKee, Bai, & Fiske, 2023).

AI systems can function as autonomous, opaque, and socially consequential agents (Daughterty & Wilson, 2024), making their adoption not only a technical decision, but also a psychological and moral one (Bonnefon et al., 2024; Ladak et al., 2024). Users must decide whether and how much they trust algorithmic decisions, whether these systems threaten their expertise or autonomy, and whether they feel capable of working effectively alongside them (Lee, Jones-Jang, Chung, Kim, & Choi, 2024). These dynamics create conditions under which affective, self-evaluative, and attitudinal factors become decisive, yet they are not adequately represented within the original UTAUT model.

To address these limitations, the present study reintroduces three key predictors that were deliberately excluded from UTAUT but central to its parent theories: attitude toward using technology, self-efficacy, and anxiety. Each construct captures a dimension of user response that is particularly salient in the context of AI adoption.

Attitude toward using technology originates from the Theory of Reasoned Action (Fishbein & Ajzen, 1975) and the Technology Acceptance Model (Davis, 1989). It reflects an individual's overall affective evaluation of using a system, including whether they find it enjoyable, valuable, or congruent with their professional self-concept. Although attitude was excluded from UTAUT because it overlapped statistically with performance and effort expectancy (Venkatesh et al., 2003), more recent research has shown that attitude contributes uniquely to explaining behavioral intention when technologies are novel, value-laden, or identity-relevant (Dwivedi et al., 2019). In the AI context, where systems can alter how employees perceive their own competence and moral responsibility, attitude captures the evaluative and identity-affirming dimensions of acceptance that expectancy beliefs alone cannot.

Self-efficacy, derived from Social Cognitive Theory (Bandura, 1986), refers to users' beliefs in their capability to perform tasks using a given technology. UTAUT subsumed self-efficacy under effort expectancy, assuming that perceived ease of use adequately captured confidence in skill execution (Venkatesh et al., 2003). Yet AI systems often require *conceptual* rather than merely procedural understanding. Specifically, users must interpret probabilistic outputs, manage uncertainty, and calibrate oversight (Daughterty & Wilson, 2024). Consequently, perceived ease may not translate into actual confidence in interacting with or supervising AI systems. Reinstating self-efficacy recognizes that confidence in one's ability to work effectively with AI is a distinct and powerful determinant of adoption (Compeau & Higgins, 1995; Marikyan, Papagiannidis, & Stewart, 2023).

Anxiety, also rooted in Social Cognitive Theory and early computer acceptance research, denotes the apprehension or emotional tension associated with technology use. Although it was initially removed from UTAUT on the grounds that its effects were inconsistent across studies (Venkatesh et al., 2003), anxiety has re-emerged as a critical construct in AI contexts. Modern AI systems often elicit anticipatory fears about job displacement, loss of professional control, or algorithmic errors—all of which can directly affect both behavioral intention and sustained engagement (Park and Jones-Jang, 2022; Jordan et al., 2008; Lee et al., 2024). Because these emotional responses operate independently of perceived usefulness or ease, reintroducing anxiety allows the model to account for the affective resistance that uniquely accompanies intelligent and autonomous systems (Venkatesh, 2021).

Together, these three constructs—attitude toward using technology, self-efficacy, and anxiety—restore the affective and self-evaluative dimensions that were stripped from UTAUT's core model but are essential for explaining adoption in AI-mediated environments. Re-embedding them within the framework offers a more holistic account of technology acceptance that integrates rational, emotional, and identity-based determinants. Moreover, Marikyan and colleagues' (2023) meta-analysis revealed that technology acceptance and use are shaped by a broader constellation of factors than those originally specified by the UTAUT. Given the rapidly evolving and deeply contextual nature of AI in today's professional workplaces, it is critical to adopt a broader theoretical lens, one that theorizes *about* context as well as *within* it, to more fully capture the multifaceted factors that shape employees' engagement with AI technologies.

Building on this extended UTAUT framework, the present study shifts attention from modeling the structural relationships among constructs to examining between-group differences in key technology acceptance variables. Rather than testing causal paths, our focus is on how employees' demographic and professional characteristics, including years of experience, organizational level, and geographic region, shape their perceptions and attitudes toward AI across eight core constructs: performance expectancy, effort expectancy, social influence, facilitating conditions, attitude toward using technology, self-efficacy, anxiety, and behavioral intention. This approach is consistent with UTAUT's original emphasis on individual differences as moderators of technology acceptance (Venkatesh et al., 2003) but extends that logic by examining systematic contextual variation across user groups. By doing so, the present analysis responds to calls for stronger contextual theorizing in technology acceptance research (Venkatesh, Thong, & Xu, 2016) and provides an empirically grounded understanding of how AI adoption dynamics differ across professional strata and organizational contexts.

**Research Questions and Hypotheses**

The present study examines how employees' professional experience, organizational level, and geographic region relate to differences in artificial intelligence (AI) adoption, usage intensity, and technology acceptance attitudes. Rather than modeling causal paths or predicting the direction of specific effects, this study employs omnibus, non-parametric tests to identify whether systematic differences exist across demographic groups on key variables derived from the Unified Theory of Acceptance and Use of Technology (UTAUT) and its extensions. This approach aligns with UTAUT's emphasis on individual and contextual moderators (Venkatesh et al., 2003) and extends it to the emerging domain of AI technologies in professional workplaces.

*Professional Experience and Attitudes Toward AI*

Years of professional experience may shape how employees perceive and engage with AI technologies. Prior UTAUT research suggests that experience moderates the influence of several key determinants, such as effort expectancy and social influence, as individuals gain familiarity and confidence in using technology (Venkatesh et al., 2003). More experienced professionals may have developed adaptive strategies and domain knowledge that influence how they interpret new technological systems, whereas newer professionals may rely more heavily on perceived ease of use or peer norms. However, the design of contemporary AI tools—emphasizing automation and usability—may attenuate these experience-related distinctions.

> **H1a:** AI adoption rates will differ across levels of professional experience.
>
> **H1b:** AI usage intensity (frequency and duration) will differ across levels of professional experience.
>
> **H1c:** UTAUT variables (performance expectancy, effort expectancy, social influence, facilitating conditions, attitude, self-efficacy, anxiety, and behavioral intention) will differ across levels of professional experience.

### *Organizational Level and Attitudes Toward AI*

Organizational level represents a critical structural factor influencing technology engagement. Senior employees typically possess greater decision-making authority, resource access, and strategic visibility—conditions that UTAUT associates with higher facilitating conditions and performance expectancy. Conversely, individuals at lower levels may encounter more task-level exposure to AI systems but less autonomy in how they are used. Additionally, social influence may vary across levels: leaders may experience normative pressure from peers and strategic expectations from the organization, whereas junior employees may experience more immediate peer or managerial influence. Given these dynamics, organizational level likely conditions both behavioral and attitudinal aspects of AI acceptance.

> **H2a:** AI adoption rates will differ across organizational levels.
>
> **H2b:** AI usage intensity (frequency and duration) will differ across organizational levels.

**H2c:** UTAUT variables will differ across organizational levels.

*Geographic Region and Attitudes Toward AI*

Geographic and cultural contexts may also shape how employees perceive and interact with AI technologies. Cross-cultural studies in technology acceptance suggest that cultural values, institutional norms, and historical experiences with technological disruption can moderate perceptions of usefulness, risk, and social influence (Im et al., 2011; Srite & Karahanna, 2006). For example, societies with higher uncertainty avoidance or lower institutional trust may exhibit stronger apprehension toward automation, whereas those with greater exposure to digital transformation may express more optimism and efficacy. Given the mixed evidence regarding cross-regional consistency in UTAUT relationships (e.g., Gupta et al., 2008; Baptista & Oliveira, 2015), this study treats regional variation as exploratory.

**H3a:** AI adoption rates will differ across geographic regions.

**H3b:** AI usage intensity (frequency and duration) will differ across geographic regions.

**H3c:** UTAUT variables will differ across geographic regions (exploratory).

Collectively, these hypotheses examine whether employees' professional experience, organizational position, and regional context are associated with statistically significant group differences in (a) AI adoption behavior, (b) AI usage intensity, and (c) UTAUT-based acceptance attitudes. This non-directional, comparative approach provides a descriptive foundation for understanding how contextual and demographic characteristics relate to emerging patterns of AI engagement in organizational settings.

**Method**

*Participants and Procedure*

Data were collected from 2,257 professionals across a single multinational consulting organization through Microsoft Forms, an online survey platform. Email invitations were distributed to all eligible participants, with reminders sent to increase response rates. The consulting context provides an appropriate setting for examining AI acceptance, as knowledge workers in this industry represent early adopters of cognitive technologies while facing direct implications of AI capabilities for their professional roles. The multinational scope enables examination of both organizational hierarchy and geographic effects within a consistent industry context.

Participants represented diverse organizational levels, geographic regions, and experience levels. Organizational levels included analyst and senior analyst, consultant and senior consultant, manager and group manager, director and senior director, and executive positions. Geographic regions encompass Europe, Growth Markets, North America, and the Global Delivery Network (GDN). Experience levels were categorized into: less than one year, 1-3 years, 3-5 years, 5-10 years, and 10 or more years. The survey was administered in English and took approximately 25 minutes to complete.

Of the total sample, 1,256 participants (55.7%) reported current use of AI tools in their work, while 1,001 (44.3%) reported no current AI tool usage. The UTAUT measures were administered exclusively to the AI user subsample for two primary reasons. First, the UTAUT framework was designed to assess attitudes and intentions among individuals who have direct experience with AI technology. Assessing UTAUT constructs among non-users would not capture meaningful technology acceptance attitudes, as these individuals lack the experiential basis for evaluating performance expectations, effort requirements, or behavioral intentions regarding actual AI usage. Second, this approach enables more precise examination of within-user variation in acceptance attitudes, as our present focus is on understanding demographic differences in AI adoption, usage behaviors, and attitudes towards the technology among active users.

*Measures*

**AI Adoption and Usage Intensity.** AI adoption was assessed with a single dichotomous item asking participants whether they used any AI tools in their regular workflow ("Do you use AI tools in your workflow?"; response options: *yes* or *no*). Among AI users, two additional items captured usage intensity. Frequency of use was measured on a six-point ordinal scale ranging from *0 = on rare occasions* to *5 = multiple times a day.* Duration of use captured the length of time respondents had incorporated AI into their workflow, rated on a six-point scale from *0 = less than one month* to *5 = over two years.* These items provided a basic behavioral profile of participants' AI adoption patterns.

**UTAUT Constructs.** Eight dimensions derived from the Unified Theory of Acceptance and Use of Technology (UTAUT) and its extensions were measured using established scales adapted to the AI context. The constructs included performance expectancy, effort expectancy, social influence, facilitating conditions, attitude toward using technology, self-efficacy, anxiety, and behavioral intention. All items were rated on seven-point Likert scales anchored at *1 = strongly disagree* and *7 = strongly agree.* Full scale items are presented in the Appendix.

**Demographic Variables.** Participants reported several demographic and professional characteristics. Years of professional experience was measured categorically, with five levels ranging from less than one year to more than ten years. Organizational level included four categories after excluding executives due to small sample size. Geographic region reflected four major business regions within the organization: Europe, Growth Markets, North America, and the Global Delivery Network. Table 1 summarizes sample distributions across these demographic groups.

**Table 1**

*Sample Distribution Across Demographic Categories*

| Demographic Variable | Category | n | % |
| --- | --- | --- | --- |
| Years of professional experience | < 1 year | 88 | 7.0 |
| | 1–3 years | 196 | 15.6 |
| | 3–5 years | 187 | 14.9 |
| | 5–10 years | 247 | 19.7 |
| | 10+ years | 533 | 42.8 |
| Organizational level | Analyst/Senior Analyst | 356 | 28.6 |
| | Consultant/Senior Consultant | 417 | 33.5 |
| | Manager/Group Manager | 372 | 29.9 |
| | Director/Senior Director | 100 | 8.0 |
| Geographic region | Europe | 558 | 44.4 |
| | Growth Markets | 348 | 27.7 |
| | North America | 323 | 25.7 |
| | Global Delivery Network | 21 | 1.7 |

*Analytic Strategy*

The analysis unfolded in four phases aligned with the study's research objectives. First, we generated descriptive statistics to summarize overall patterns of AI adoption and usage.

Second, we examined whether demographic characteristics predicted AI adoption status (user vs. non-user) using chi-square tests of independence. This nonparametric approach was appropriate given the categorical nature of both predictor variables (demographic categories) and the outcome variable. To meet chi-square assumptions requiring expected frequencies of at least five per cell (Cochran, 1954), we excluded participants at the executive organizational level (n = 7) due to insufficient cell counts. We quantified effect sizes using Cramer's V, with values of .01, .06, and .14 interpreted as small, medium, and large effects, respectively (Cohen, 1988).

Third, among AI users only (n = 1,256), we assessed the internal consistency reliability of each UTAUT scale using Cronbach's alpha coefficients, interpreting values of α ≥ .70 as acceptable (Nunnally & Bernstein, 1994). We also examined bivariate correlations among the UTAUT dimensions using Pearson correlation coefficients with significance testing. In addition, we analyzed usage intensity, measured as self-reported frequency and duration of AI tool use, using Kruskal–Wallis H tests. This nonparametric alternative to one-way ANOVA was appropriate given the ordinal nature of the data and violations of normality observed in preliminary analyses (Kruskal & Wallis, 1952).

Fourth, before comparing UTAUT attitudes across demographic groups, we systematically tested the assumptions of parametric analyses. We evaluated normality using Shapiro–Wilk tests (Shapiro & Wilk, 1965) and homogeneity of variance using Levene's tests (Levene, 1960). Significant violations of normality across several UTAUT variables prompted

EMPLOYEE AI ADOPTION AND USAGE                                                                              21the continued use of nonparametric methods. Accordingly, we conducted Kruskal–Wallis H tests to compare UTAUT dimensions across demographic categories, applying a Bonferroni correction to control the familywise error rate ($\alpha = .05 / 8 = .006$). We reported epsilon-squared ($\varepsilon^2$) to estimate effect sizes for these comparisons (Tomczak & Tomczak, 2014).

## Results

### AI Adoption

We first examined whether AI adoption differed across years of experience, organizational level, and geographic region (H1a, H2a, and H3a).

**Years of Experience.** Years of professional experience showed a marginally non-significant association with AI adoption, $\chi^2(4, N = 2{,}257) = 8.11$, $p = .088$, Cramer's V = .060 (see Table 2). Thus, H1a was not supported.

**Table 2**

*AI Usage Patterns by Years of Experience*

| Years of Experience | Total N | AI Non-Users (n) | AI Users (n) | Usage Rate (%) |
|---|---|---|---|---|
| Less than one year | 142 | 54 | 88 | 62.0% |
| 1-3 years | 365 | 169 | 196 | 53.7% |
| 3-5 years | 310 | 122 | 188 | 60.6% |
| 5-10 years | 436 | 189 | 247 | 56.7% |
| 10+ years | 997 | 464 | 533 | 53.5% |

**Organizational Level.** Chi-square analysis revealed a significant association between organizational level and AI adoption, $\chi^2(3, N = 2{,}241) = 25.42$, $p < .001$, Cramer's V = .106. As shown in Table 3, adoption rates ranged from 50.0% among consultants/senior consultants to 70.4% among directors/senior directors. Although the effect size was small, the result indicates that adoption rates are not uniform across levels, supporting H2a.

**Table 3**

*AI Usage Patterns by Organizational Level*

| Organizational Level | Total N | AI Non-Users (n) | AI Users (n) | Usage Rate (%) |
|---|---|---|---|---|
| Analyst/Sr Analyst | 620 | 264 | 356 | 57.4% |
| Consultant/Sr Consultant | 836 | 418 | 418 | 50.0% |
| Manager/Group Manager | 643 | 271 | 372 | 57.9% |
| Director/Sr Director | 142 | 42 | 100 | 70.4% |

Note: Executive level excluded due to small sample size (n < 10).

**Geographic Region.** Adoption rates were statistically similar across regions, $\chi^2(3, N = 2{,}257) = 1.02$, $p = .797$, Cramer's V = .021 (see Table 4). Accordingly, H3a was not supported.

**Table 4**

*AI Usage Patterns by Region*

| Region | Total N | AI Non-Users (n) | AI Users (n) | Usage Rate (%) |
|---|---|---|---|---|
| Europe | 1016 | 458 | 558 | 54.9% |
| GDN | 37 | 16 | 21 | 56.8% |
| Growth Markets | 629 | 281 | 348 | 55.3% |
| North America | 562 | 239 | 323 | 57.5% |

Overall, organizational level emerged as the only demographic variable significantly associated with AI adoption, whereas experience and regional differences were negligible.

*AI Usage Intensity Among AI Users*

Next, we tested whether frequency and duration of AI use differed across demographic groups (H1b, H2b, H3b), focusing on only AI users ($n = 1{,}256$).

**Usage Frequency.** As seen in Table 5, no significant differences emerged for organizational level ($H(3) = 3.60$, $p = .309$, $\varepsilon^2 = .001$), years of experience ($H(4) = 0.78$, $p = .941$, $\varepsilon^2 < .001$), or geographic region ($H(3) = 2.66$, $p = .448$, $\varepsilon^2 < .001$).

**Usage Duration.** Duration of AI use showed small omnibus differences by organizational level ($H(3) = 10.32$, $p = .016$, $\varepsilon^2 = .006$) and experience ($H(4) = 16.47$, $p = .002$, $\varepsilon^2 = .010$), but neither survived Bonferroni correction ($\alpha = .006$). Regional differences were nonsignificant ($H(3) = 4.16$, $p = .245$, $\varepsilon^2 = .001$).

Given the small effect sizes observed, these results provide only limited evidence for H1b, H2b, and H3b, suggesting that once employees adopt AI, their usage frequency and duration are broadly similar across demographic groups.

**Table 5**

*AI Usage Intensity Across Demographic Groups*

| Demographic Variable | Usage Measure | Groups (n) | M (SD) | H | df | p | $\varepsilon^2$ |
|---|---|---|---|---|---|---|---|
| Organizational Level | Frequency | Analyst (355) | 4.24 (1.13) | 3.596 | 3 | .309 | .001 |
| | | Consultant (418) | 4.29 (1.11) | | | | |
| | | Manager (372) | 4.24 (1.11) | | | | |
| | | Director (100) | 4.50 (0.70) | | | | |
| | Duration | Analyst (356) | 2.36 (1.25) | 10.320 | 3 | *.016* | .006 |
| | | Consultant (418) | 2.33 (1.17) | | | | |
| | | Manager (372) | 2.50 (1.24) | | | | |
| | | Director (100) | 2.72 (1.07) | | | | |
| Years of Experience | Frequency | < 1 year (88) | 4.36 (1.01) | 0.784 | 4 | .941 | −.003 |
| | | 1–3 years (195) | 4.24 (1.14) | | | | |
| | | 3–5 years (188) | 4.27 (1.12) | | | | |
| | | 5–10 years (247) | 4.23 (1.15) | | | | |
| | | 10+ years (533) | 4.32 (1.04) | | | | |
| | Duration | < 1 year (88) | 2.05 (1.35) | 16.466 | 4 | .002 | .010 |
| | | 1–3 years (196) | 2.36 (1.21) | | | | |
| | | 3–5 years (188) | 2.45 (1.18) | | | | |
| | | 5–10 years (247) | 2.29 (1.24) | | | | |
| | | 10+ years (533) | 2.56 (1.17) | | | | |
| Geographic Region | Frequency | North America (323) | 4.22 (1.13) | 2.657 | 3 | .448 | −.0003 |
| | | Europe (557) | 4.29 (1.06) | | | | |
| | | Growth Markets (348) | 4.32 (1.10) | | | | |
| | | GDN (21) | 4.29 (1.08) | | | | |
| | Duration | North America (323) | 2.44 (1.21) | 4.155 | 3 | .245 | .001 |
| | | Europe (558) | 2.50 (1.19) | | | | |
| | | Growth Markets (348) | 2.31 (1.23) | | | | |
| | | GDN (21) | 2.24 (1.51) | | | | |

Note: N = 1,256 AI adopters. Kruskal-Wallis tests used due to non-normal distributions. Executive group excluded from organizational analysis (n = 7). *p < .05. **p < .01. ***p < .0001.

*UTAUT Variables*

**Reliability and Correlation.** Internal consistency analysis revealed that seven of eight UTAUT variables demonstrated acceptable to excellent reliability. Five constructs achieved strong reliability: performance expectancy ($\alpha = 0.81$), effort expectancy ($\alpha = 0.89$), attitude toward AI ($\alpha = 0.86$), self-efficacy ($\alpha = 0.84$), and anxiety ($\alpha = 0.87$). Two additional constructs demonstrated acceptable reliability: social influence ($\alpha = 0.79$) and facilitating conditions ($\alpha = 0.73$). However, behavioral intention exhibited questionable internal consistency ($\alpha = 0.63$), suggesting some challenges with scale coherence within this construct and warranting cautious interpretation of findings involving this measure.

Correlational analyses revealed theoretically consistent patterns among UTAUT constructs (see Figure 3). Performance expectancy demonstrated strong positive associations with attitude toward AI ($r = 0.64$, $p < .001$) and moderate positive association with behavioral intention ($r = 0.30$, $p < .001$), consistent with UTAUT theoretical predictions regarding the central role of performance beliefs in technology acceptance. Anxiety exhibited inverse relationships with most other UTAUT dimensions, supporting its conceptualization as a barrier to technology acceptance.

**Figure 3**

*UTAUT Correlations (AI Users Only)*

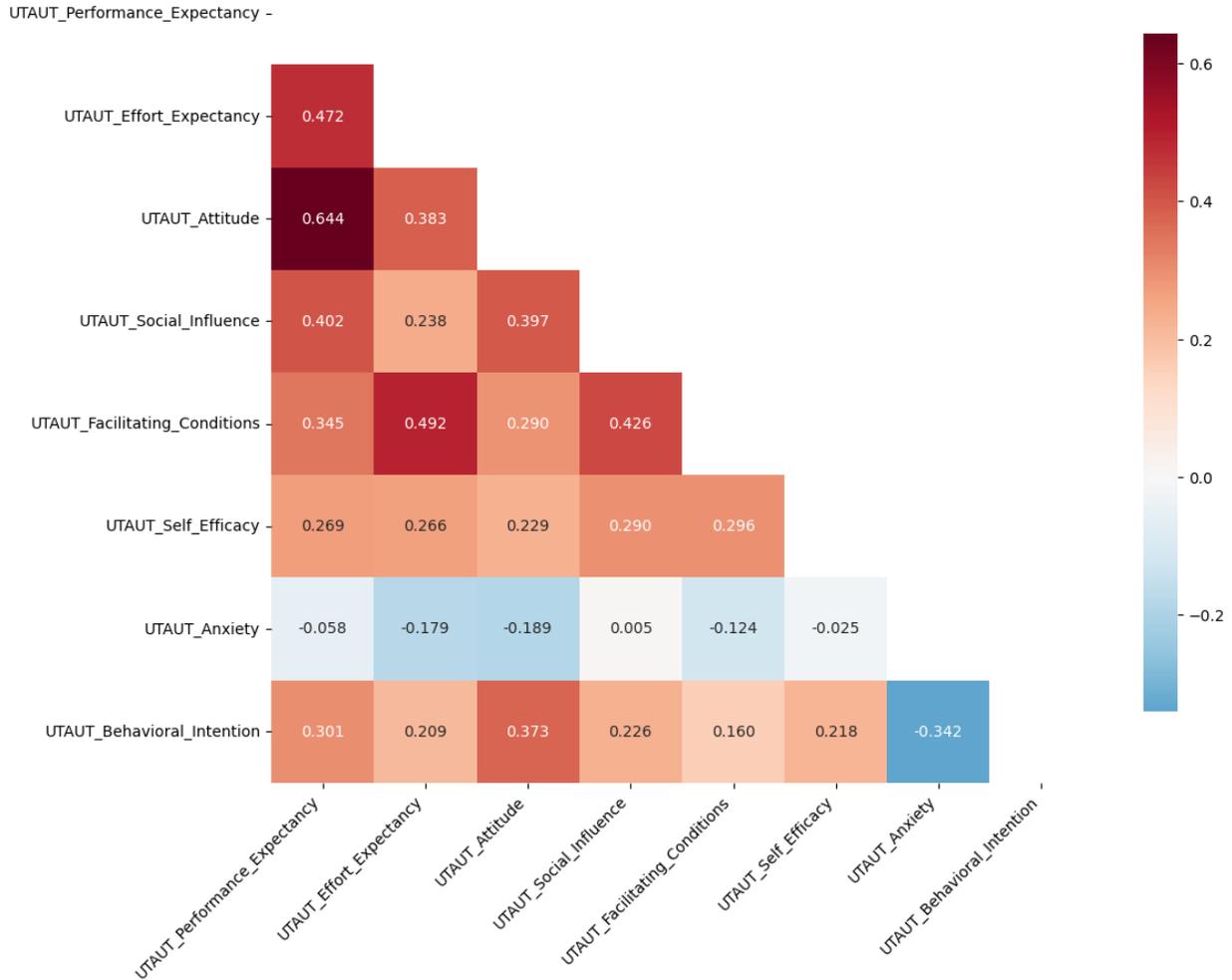

**Demographic Differences.** To examine whether technology-acceptance attitudes differed across key demographic and contextual factors, we conducted Kruskal–Wallis tests to compare UTAUT composite scores across years of professional experience, organizational level, and geographic region among AI users ($n = 1,256$). Because these analyses represent omnibus, non-parametric tests, they identify whether significant differences exist among groups but do not specify which specific groups differ from one another. In line with the study's focus on overall demographic variation rather than detailed pairwise contrasts, post-hoc tests were not conducted. Bonferroni adjustment was applied to control the familywise error rate across eight dependent variables, resulting in a conservative significance threshold of α = .006. All results are reported in Table 6.

**Table 6**

*Demographic Differences in UTAUT Variables Based on Kruskal–Wallis Analyses*

| UTAUT Variable | Demographic | H | df | p | ε² | Effect | Sig |
|---|---|---|---|---|---|---|---|
| Anxiety | Experience | 44.62 | 4 | <.001 | 0.033 | small | *** |
|  | Org Level | 44.24 | 3 | <.001 | 0.033 | small | *** |
|  | Region | 18.71 | 3 | <.001 | 0.013 | small | *** |
| Attitude | Experience | 1.17 | 4 | 0.884 | 0.000 | negligible | ns |
|  | Org Level | 1.82 | 3 | 0.610 | 0.000 | negligible | ns |
|  | Region | 27.40 | 3 | <.001 | 0.020 | small | *** |
| Behavioral Intention | Experience | 14.65 | 4 | 0.005 | 0.009 | negligible | ** |
|  | Org Level | 28.36 | 3 | <.001 | 0.021 | small | *** |
|  | Region | 10.31 | 3 | 0.016 | 0.006 | negligible | ns |
| Effort Expectancy | Experience | 14.80 | 4 | 0.005 | 0.009 | negligible | ** |
|  | Org Level | 30.23 | 3 | <.001 | 0.022 | small | *** |
|  | Region | 2.94 | 3 | 0.401 | 0.000 | negligible | ns |
| Facilitating Conditions | Experience | 5.94 | 4 | 0.204 | 0.002 | negligible | ns |
|  | Org Level | 6.98 | 3 | 0.073 | 0.003 | negligible | ns |
|  | Region | 10.61 | 3 | 0.014 | 0.006 | negligible | ns |
| Performance Expectancy | Experience | 11.83 | 4 | 0.019 | 0.006 | negligible | ns |
|  | Org Level | 22.73 | 3 | <.001 | 0.016 | small | *** |
|  | Region | 38.91 | 3 | <.001 | 0.029 | small | *** |
| Self Efficacy | Experience | 1.87 | 4 | 0.759 | 0.000 | negligible | ns |
|  | Org Level | 1.93 | 3 | 0.587 | 0.000 | negligible | ns |
|  | Region | 18.02 | 3 | <.001 | 0.012 | small | *** |
| Social Influence | Experience | 9.28 | 4 | 0.054 | 0.004 | negligible | ns |
|  | Org Level | 6.24 | 3 | 0.101 | 0.003 | negligible | ns |
|  | Region | 54.83 | 3 | <.001 | 0.042 | small | *** |

***Years of Experience Effects.*** The analyses revealed significant omnibus differences across years of professional experience for three of the eight UTAUT constructs: anxiety, effort expectancy, and behavioral intention. Specifically, anxiety demonstrated the largest effect, $H(4) = 44.62$, $p < .001$, $ε² = .033$, indicating that employees' emotional responses to AI varied meaningfully by experience level. Effort expectancy also differed significantly across groups, $H(4) = 14.81$, p = .005, $\varepsilon^2 = .009$, although the effect size was minimal and likely reflects small perceptual variations in the ease of using AI technologies. Behavioral intention likewise showed significant but modest omnibus differences, $H(4) = 14.65$, p = .006, $\varepsilon^2 = .008$.

Together, these results suggest that professional experience may play a role in shaping certain aspects of AI acceptance, particularly affective and motivational dimensions, while the remaining five constructs (performance expectancy, attitude, self-efficacy, social influence, and facilitating conditions) were statistically equivalent across experience groups after correction for multiple comparisons. Because the Kruskal–Wallis test does not reveal where these differences lie, further post-hoc testing (e.g., Dunn's pairwise comparisons with Bonferroni adjustment) would be necessary to determine which specific experience levels differ significantly from one another.

***Organizational Level Effects.*** Significant omnibus differences were also observed across organizational levels for four UTAUT constructs: anxiety, effort expectancy, performance expectancy, and behavioral intention. Anxiety again exhibited the strongest effect, $H(3) = 44.24$, $p < .001$, $\varepsilon^2 = .033$, suggesting that employees' emotional responses toward AI differ by hierarchical position within an organization. Effort expectancy, $H(3) = 30.23$, $p < .001$, $\varepsilon^2 = .022$, and performance expectancy, $H(3) = 22.73$, $p < .001$, $\varepsilon^2 = .016$, also varied across levels, indicating that perceptions of ease of use and performance benefit are not uniform throughout the organizational hierarchy.

Finally, behavioral intention demonstrated significant omnibus variation, $H(3) = 28.36$, $p < .001$, $\varepsilon^2 = .020$, implying that employees at different levels may differ in their general willingness or motivation to use AI systems. The remaining four constructs—attitude toward AI, social influence, facilitating conditions, and self-efficacy—did not differ significantly across levels. Because Kruskal–Wallis tests identify only the presence of significant group differences, these findings should be interpreted as indicating that organizational level is associated with some heterogeneity in AI-related perceptions, though the specific nature and direction of those differences remain undetermined without post-hoc comparisons.

*Regional Effects.* Geographic region yielded the broadest pattern of omnibus differences among the UTAUT constructs, with five of eight dimensions showing statistically significant variation across regions. Significant effects were observed for anxiety, $H(3) = 18.71, p < .001, \varepsilon^2 = .013$; performance expectancy, $H(3) = 38.91, p < .001, \varepsilon^2 = .029$; attitude toward AI, $H(3) = 27.40, p < .001, \varepsilon^2 = .020$; social influence, $H(3) = 54.83, p < .001, \varepsilon^2 = .042$; and self-efficacy, $H(3) = 18.02, p < .001, \varepsilon^2 = .012$. These results indicate that employees' perceptions of AI's usefulness, their affective evaluations, and their confidence and perceived social expectations surrounding AI use vary across global regions. However, because the Kruskal–Wallis test provides only an omnibus assessment, these findings do not identify which specific regional pairs differ (e.g., North America vs. Europe). The remaining three constructs—effort expectancy, facilitating conditions, and behavioral intention—did not differ significantly across regions.

Overall, the regional analyses support the presence of contextual variation in AI-related attitudes, but the magnitude of these effects was modest, and further post-hoc analyses would be needed to map the specific sources of cross-regional difference.

**Summary.** Across demographic factors, statistically significant omnibus effects emerged for several UTAUT constructs, but all effect sizes were small ($\varepsilon^2 = .008–.041$). These results indicate that demographic and contextual characteristics are associated with some variation in technology-acceptance attitudes, particularly in emotional (anxiety), cognitive (performance and effort expectancy), and motivational (behavioral intention) domains. However, because the present analyses identify only the existence of between-group differences rather than their specific pairwise patterns, these findings should be interpreted as preliminary evidence of heterogeneity rather than as evidence of systematic directional trends. Additional post-hoc tests would be necessary to determine precisely which groups differ from one another and to characterize the substantive meaning of those differences.

**Discussion**

The present study examined demographic and contextual variation in employees' adoption and acceptance of AI technologies within the workplace, drawing on an extended Unified Theory of Acceptance and Use of Technology (UTAUT) framework that incorporates both cognitive and affective dimensions of technology acceptance. Rather than testing structural paths among constructs, the study focused on whether employees' professional experience, organizational level, and geographic region were associated with differences in adoption behavior, AI usage intensity, and key UTAUT variables.

Using non-parametric omnibus tests, we found that although demographic characteristics accounted for statistically significant variation across several constructs, these effects were small in magnitude. Importantly, the findings suggest that while organizational position and regional context correspond to modest differences in how employees perceive and engage with AI systems, much of the variance in technology acceptance likely reflects individual-level psychological processes rather than demographic attributes.

*Demographic Patterns in AI Adoption and Use*

Consistent with expectations, organizational level emerged as a strong demographic correlate of AI adoption. Employees in more senior roles reported higher adoption rates than those in junior or mid-level positions, suggesting that hierarchical position confers greater opportunity, discretion, or incentive to engage with AI tools. This finding aligns with UTAUT's emphasis on facilitating conditions such as access to resources, autonomy, and strategic visibility as critical enablers of technology use (Venkatesh et al., 2003). It also resonates with organizational research highlighting that decision-making authority and perceived control can increase willingness to adopt new technologies, even when the perceived risks or uncertainties are high (Dwivedi et al., 2019).

By contrast, years of experience and geographic region were not significantly related to AI adoption, indicating that exposure to or familiarity with prior technologies does not necessarily predict whether employees integrate AI into their workflow. Similarly, usage frequency and duration were largely consistent across all demographic groups. These findings suggest that once employees begin using AI tools, patterns of continued use become relatively uniform, reflecting the integration of AI into routine work processes rather than discretionary variation based on background characteristics.

Together, these results imply that demographic factors may be more relevant to initial adoption decisions than to ongoing engagement, supporting recent evidence that technology acceptance transitions from being socially and demographically driven in early stages to habitually and psychologically driven once use is established (Marikyan et al., 2023; Venkatesh et al., 2016).

### *Demographic Differences in Technology Acceptance Attitudes*

Beyond behavioral adoption, several UTAUT constructs differed significantly across demographic categories, though again with small effect sizes. These findings suggest that contextual differences exist in how employees think and feel about AI, even if such differences do not strongly translate into behavioral variation. Professional experience was associated with differences in anxiety, effort expectancy, and behavioral intention. Organizational level corresponded with differences in anxiety, effort expectancy, performance expectancy, and behavioral intention, while geographic region showed the broadest set of omnibus effects, spanning anxiety, performance expectancy, attitude, social influence, and self-efficacy. Importantly, these tests identify that groups are not equivalent but do not specify which pairs differ; thus, interpretations regarding the direction or substantive pattern of effects remain unknown.

      The recurring appearance of AI anxiety as a significant construct across experience, level, and region underscores its centrality in the acceptance of AI technologies. Anxiety has long been recognized as a salient, albeit under-theorized, determinant of technology adoption (Venkatesh, 2021). In the AI context, it reflects not merely apprehension about technical proficiency but broader concerns about autonomy, professional identity, and algorithmic control (Lee et al., 2024; McKee et al., 2023). That anxiety varied across all three demographic variables suggests that emotional responses to AI are contextually embedded and shaped by employees' experience base, hierarchical position, and cultural environment. However, because the Kruskal–Wallis tests identify only the existence of differences and not their specific direction, future research using targeted post-hoc or multilevel modeling approaches is needed to clarify how these contextual features amplify or mitigate anxiety in concrete ways.

      Variation in performance expectancy and effort expectancy likewise suggests that perceptions of AI's usefulness and ease of use are not entirely uniform across employees. These findings are consistent with UTAUT's prediction that the salience of these cognitive constructs may vary with experience and control (Venkatesh et al., 2003; Dwivedi et al., 2019). However, the small effect sizes observed here suggest that modern AI tools may be sufficiently intuitive

that perceived ease of use no longer serves as a meaningful differentiator, a pattern noted in previous research on mature technologies (Gupta et al., 2008). The observed differences in behavioral intention mirror this pattern, implying that contextual factors exert subtle but discernible influences on employees' motivational readiness to use AI, even as actual usage behaviors remain relatively consistent.

Finally, the finding that regional differences emerged for multiple constructs, including social influence and self-efficacy, highlights the potential role of cultural norms, communication patterns, and local attitudes toward automation in shaping AI acceptance. Although UTAUT does not explicitly theorize cultural variability, prior cross-national research has shown that collectivistic and uncertainty-avoidant cultures may place greater emphasis on social endorsement and perceived control in technology adoption (Im et al., 2011; Srite & Karahanna, 2006). The current results lend preliminary support to this possibility within a global organizational context, suggesting that cultural and institutional environments may subtly condition the weight employees assign to social and efficacy-based considerations when evaluating whether and how to use AI technologies.

## *Theoretical Implications*

The findings offer several theoretical contributions to the literature on technology acceptance and organizational behavior. First, they extend UTAUT by reaffirming the importance of contextual moderators like region and culture, an element often acknowledged but rarely tested empirically. By examining demographic and organizational factors as sources of variability in UTAUT constructs, this study complements prior research that has focused primarily on individual-level predictors (Venkatesh et al., 2016). Second, the inclusion of affective and self-evaluative constructs, particularly anxiety and self-efficacy, demonstrates the

EMPLOYEE AI ADOPTION AND USAGE																																				34value of revisiting elements excluded from the original UTAUT but central to its parent theories. The consistent significance of anxiety across analyses suggests that purely cognitive accounts of technology acceptance may underestimate the affective and identity-related dimensions of AI adoption.

More broadly, the results underscore the need for contextual theorizing about AI acceptance in organizational settings. As Venkatesh and colleagues (2016) argue, applying UTAUT "in context" is not the same as theorizing "about context." The current findings illustrate that demographic and organizational variables, while statistically modest in effect, meaningfully shape the affective and perceptual contours of technology acceptance. Incorporating these contextual dynamics into theory building can help explain why identical technologies produce divergent experiences across professional strata and cultural environments.

*Practical Implications*

From a practical standpoint, the results suggest that efforts to foster AI adoption and comfort should prioritize organizational-level and affective interventions rather than focusing narrowly on technical training or user experience design. Because seniority was the strongest correlate of adoption, organizations may benefit from explicitly supporting middle and junior employees, who may have less discretion and fewer resources, to integrate AI meaningfully into their workflows. Leadership modeling and transparent communication about the strategic purpose of AI may also reduce uncertainty and anxiety among less senior employees. Additionally, given the small but consistent associations between demographic variables and anxiety, interventions that normalize learning, emphasize psychological safety, and frame AI as augmentative rather than substitutive could further encourage equitable engagement across employee groups.

The limited regional differences observed suggest that global organizations can pursue largely standardized implementation strategies, provided they remain sensitive to local communication norms and cultural variations in perceived control and social endorsement. Collectively, these insights point to the importance of combining structural supports (e.g., access, autonomy, training) with cultural and emotional enablers (e.g., trust, efficacy, belonging) to promote sustainable, organization-wide AI adoption.

### *Limitations and Directions for Future Research*

Several limitations temper the conclusions drawn from this study. Most notably, the use of Kruskal–Wallis tests limits interpretation to omnibus effects, indicating only that group differences exist somewhere among categories without specifying which pairs differ. Post-hoc analyses (e.g., Dunn's test or Bonferroni-adjusted pairwise comparisons) were not conducted, given the study's focus on whether demographic variation exists at all rather than delineating specific contrasts. Future research should employ such analyses, as well as hierarchical or multigroup structural modeling, to identify where and why differences occur.

In addition, the cross-sectional design precludes causal inference and cannot capture how attitudes and usage patterns evolve over time. Longitudinal designs could examine whether initial demographic differences in adoption converge as AI use becomes normalized. Finally, the study's single-organizational context may limit generalizability, as workplace norms, industry demands, and organizational culture can influence both technology adoption and the salience of UTAUT constructs. Expanding future research to multiple organizations and industries would help establish the boundary conditions of these findings.

### *Conclusion*

      Taken together, the results of this study indicate that demographic and contextual characteristics, particularly organizational level and geographic region, correspond to small but consistent differences in AI-related perceptions and attitudes. However, these differences account for limited variance relative to individual-level psychological determinants. The extended UTAUT framework applied here demonstrates that affective constructs such as anxiety and self-efficacy remain critical to understanding AI acceptance, particularly as organizations integrate increasingly autonomous and socially complex technologies into daily work. By highlighting where context matters and where it does not, this study contributes to a more nuanced understanding of technology acceptance in modern workplaces and underscores the need for both organizational and emotional infrastructures that support equitable and confident engagement with AI.

EMPLOYEE AI ADOPTION AND USAGE                                                                40

EMPLOYEE AI ADOPTION AND USAGE                                                                    42Park, Y. J., & Jones-Jang, S. M. (2022). Surveillance, privacy, and algorithmic inequality: A study of workers' perspectives. *Journal of Communication, 72*(6), 680-695. https://doi.org/10.1093/joc/jqac031

Pew Research Center. (2025, February 25). *U.S. workers are more worried than hopeful about future AI use in the workplace*. https://www.pewresearch.org/social-trends/2025/02/25/u-s-workers-are-more-worried-than-hopeful-about-future-ai-use-in-the-workplace/

Qlik. (2024). *State of data literacy report 2024*. https://www.qlik.com/us/data-literacy

Shapiro, S. S., & Wilk, M. B. (1965). An analysis of variance test for normality (complete samples). *Biometrika, 52*(3-4), 591-611. https://doi.org/10.1093/biomet/52.3-4.591

Srite, M., & Karahanna, E. (2006). The role of espoused national cultural values in technology acceptance. *MIS Quarterly, 30*(3), 679-704. https://doi.org/10.2307/25148745

Stanford University. (2024). *The 2024 AI Index Report*. Stanford Institute for Human-Centered Artificial Intelligence. https://aiindex.stanford.edu/report/

Sun, X., Xu, W., & Xu, Y. (2024). Understanding AI acceptance in the workplace: The role of trust and anxiety. *International Journal of Human-Computer Interaction, 40*(5), 1245-1260. https://doi.org/10.1080/10447318.2023.2189765

Taiwo, A. A., & Downe, A. G. (2013). The theory of user acceptance and use of technology (UTAUT): A meta-analytic review of empirical findings. *Journal of Theoretical & Applied Information Technology, 49*(1), 48-58.

Tomczak, M., & Tomczak, E. (2014). The need to report effect size estimates revisited: An overview of some recommended measures of effect size. *Trends in Sport Sciences, 21*(1), 19-25.

EMPLOYEE AI ADOPTION AND USAGE                                                              43Venkatesh, V. (2021). Adoption and use of AI tools: A research agenda grounded in UTAUT. *Annals of Operations Research*, 1-20. https://doi.org/10.1007/s10479-020-03918-9

Venkatesh, V., & Davis, F. D. (2000). A theoretical extension of the technology acceptance model: Four longitudinal field studies. *Management Science, 46*(2), 186-204. https://doi.org/10.1287/mnsc.46.2.186.11926

Venkatesh, V., Morris, M. G., Davis, G. B., & Davis, F. D. (2003). User acceptance of information technology: Toward a unified view. *MIS Quarterly, 27*(3), 425-478. https://doi.org/10.2307/30036540

Venkatesh, V., Thong, J. Y., & Xu, X. (2012). Consumer acceptance and use of information technology: Extending the unified theory of acceptance and use of technology. *MIS Quarterly, 36*(1), 157-178. https://doi.org/10.2307/41410412

Venkatesh, V., Thong, J. Y. L., & Xu, X. (2016). Unified theory of acceptance and use of technology: A synthesis and the road ahead. *Journal of the Association for Information Systems, 17*(5), 328-376. https://doi.org/10.17705/1jais.00428

Williams, M. D., Rana, N. P., & Dwivedi, Y. K. (2015). The unified theory of acceptance and use of technology (UTAUT): A literature review. *Journal of Enterprise Information Management, 28*(3), 443-488. https://doi.org/10.1108/JEIM-09-2014-0088

Zhou, T., Lu, Y., & Wang, B. (2010). Integrating TTF and UTAUT to explain mobile banking user adoption. *Computers in Human Behavior, 26*(4), 760-767. https://doi.org/10.1016/j.chb.2010.01.013

# Appendix

*Unified Theory of Acceptance and Use of Technology (UTAUT) Scale Items, Adapted for AI-Mediated Work Environment*

Performance Expectancy

1. I find AI useful in my job.

2. Using AI enables me to accomplish tasks more quickly.

3. Using AI increases my productivity.

    If I use AI, I will increase my chances of getting a raise.

Effort Expectancy

5. My interaction with AI is clear and understandable.

6. It is easy for me to become skillful at using AI.

7. I find AI easy to use.

8. Learning to operate AI is easy for me.

Attitude Toward Using Technology

9. Using AI is a good idea.

10. AI makes work more interesting.

11. I like working with AI.

Social Influence

12. People who influence my behavior think that I should use AI.

13. People who are important to me think that I should use AI.

14. Senior management has promoted the use of AI.

15. In general, my organization has supported the use of AI.

Facilitating Conditions

16. I have the resources necessary to use AI.

17. I have the knowledge necessary to use AI.

18. AI is compatible with other systems I use.

19. A specific person (or group) is available for assistance with AI difficulties.

Self-Efficacy

20. I could complete a job or task using AI if there was no one around to tell me what to do as I go.

21. I could complete a job or task using AI if I could call someone for help if I got stuck.

22. I could complete a job or task using AI if I had a lot of time to complete the job for which the software was provided.

23. I could complete a job or task using AI if I had assistance.

Anxiety

24. I feel apprehensive about using AI.

25. It scares me to think that I could lose a lot of information using AI by hitting the wrong key.

26. I hesitate to use AI for fear of making mistakes I cannot correct.

27. AI is somewhat intimidating to me.

Behavioral Intention to Use

28. I intend to use AI in the next month.

29. I intend to use AI in the next six months.

30. I intend to use AI in the next year.

31. I do not intend to use AI in the future.